\documentstyle[prl,aps,multicol,epsfig]{revtex}
\begin{document}
\draft
\tighten

\title{ Spontaneous orientation of a quantum lattice string.}
\author{Osman Y. Osman, Wim van Saarloos and Jan Zaanen}
\address{ Instituut-Lorentz, Leiden University, P.O.B. 9506, 2300 RA Leiden,
The Netherlands}
\date{\today ; osman@lorentz.leidenuniv.nl}
\maketitle

\begin{abstract}
Using exact diagonalization and quantum Monte-Carlo techniques we study 
a quantum lattice string model introduced as a model for a single cuprate 
stripe. We focus on the ground state properties of the string. Our 
result shows that in the physically relevant region of the parameter space 
a zero-temperature spontaneous symmetry breaking occurs. The string 
spontaneously orients itself along one direction in space and becomes 
directed. We introduce an order parameter for the directedness and show 
that at zero temperature this order parameter reaches its saturation value. 
\end{abstract}

\begin{multicols}{2}
\narrowtext

Since the experimental discovery of the stripe phase\cite{tran}, interest in 
this field has grown rapidly. Many issues concerning stripes are discussed, 
ranging from their origin to their relation to high $T_c$ superconductivity, 
including the dynamical properties of the stripes. In this contribution we are 
concerned with the last subject. We focus on the problem of a single stripe/ 
single charged domain wall. We consider the domain wall to be a connected 
trajectory (string) of particles, communicating with the lattice, while the 
precise nature of these particles is not further specified: the quantum 
lattice string (QLS) model\cite{eskes1}. Studying this model numerically, we 
discovered a zero-temperature symmetry breaking: although the 
string can be quantum delocalized, it picks spontaneously a {\em
  direction} in space. This symmetry breaking happens always in the
part of parameter space which is of physical relevance. 
At first sight, one might expect that the quantum fluctuations (kinetic 
energy) would tend to disorder the string, i.e., to decrease the tendency 
for the string to be directed. That the opposite effect happens, can be seen 
as follows. A first intuition can be obtained by considering the analogy with 
surface statistical mechanics. The quantum string problem can be formulated 
as a classical problem of a two dimensional surface (world sheet) in 2+1 
dimension, where the third direction is the imaginary time direction. The 
larger the kinetic term, or the smaller the temperature, the further the 
worldsheet stretches out in the time direction. At zero temperature, the
worldsheet becomes infinite in this direction as well. The statistical
physics of a string is then equivalent to that of a fluctuating sheet
in three dimensions. Now, it is well known from studies of classical
interfaces\cite{wee77} that while a one-dimensional classical
interface in two dimensions does not stay directed due to the strong
fluctuations, for a two-dimensional sheet the entropic fluctuations
are so small that interfaces can stay macroscopically flat in the
presence of a lattice\cite{Wee80,Bei87}. In other words, even if microscopic
configurations with overhangs are allowed, a classical interface on a
lattice in three dimensions can stay macroscopically flat or ``directed''. 
In the present context, we will show that the directedness is a caused by 
an order-out-of disorder mechanism: in order to maximize the fluctuations 
transversal to the local string directions, overhangs should be avoided on 
the worldsheet. It remains to be seen if this mechanism is of a more general 
application. 

In the QLS model the string configurations are specified by the position of 
the particles ${\bf r}_l = (x_{l},y_{l})$. Two consecutive particles $l$ 
and $l+1$ should either be nearest or next-nearest neighbors, i.e. $|{\bf
  r}_{l+1} - {\bf r}_l| = 1$ or $\sqrt{2}$. The set of all such configurations 
is the string Hilbert space. 

\begin{figure}
\hspace{7ex}\epsfig{figure=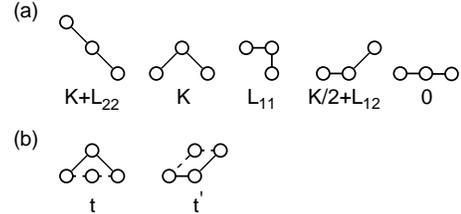,width=6cm}   
\caption{{\footnotesize (a) The set of local configurations and their 
classical energies. (b) The two allowed hoppings. We take  
$t = t^\prime$. }}
\label{fig_energies}
\end{figure}

 The Hamiltonian consists of a classical energy term and a quantum (hopping) 
term. The classical energy is a sum of local interactions between nearest 
and next nearest particles in the string. 
\vskip -0.5cm
\begin{eqnarray}
  {\cal H}_{Cl} = & & \sum_l \left[ {\cal K} \delta( | x_{l+1} - x_l | -
  1 ) \delta( | y_{l+1} - y_l | - 1 ) \right.  \nonumber \\ +
  \sum\limits_{i,j = 1}^{2} & &  \left. {\cal L}_{ij} \delta( | x_{l+1} -
  x_{l-1} | - i) \delta( | y_{l+1} - y_{l-1} | - j ) \right]~.  \label{HCl}
\end{eqnarray}
\vskip -0.2cm
\noindent with ${\cal L}_{ij} = {\cal L}_{ji}$ (see Fig.\ \ref{fig_energies})

The quantum term allows the particles to hop to nearest neighbor lattice 
positions, giving rise to the meandering of the whole string. These hops  
should respect the string constraint. To enforce the constraint a projection 
operator $P_{Str}({\bf r}) = \delta( |{\bf r}| - 1 ) + \delta( |{\bf r}| - 
\sqrt{2} )$ is introduced which insures that the motion of particle 
$l$ keeps the string intact. The string is quantized by introducing 
conjugate momenta $\pi^{\alpha}_l$, $[ r^{\alpha}_l, \pi^{\beta}_{m} ] = i 
\delta_{l,m} \delta_{\alpha,\beta}$, and the hopping is described by $e^{ i 
\pi^{x}_l} | x_{l} \rangle = | x_{l} + 1 \rangle.$ The kinetic energy becomes, 
(Fig.\ \ref{fig_energies})
\begin{equation}
  {\cal H}_{Q} = 2{\cal T} \sum\limits_{l,\alpha} P^\alpha_{Str}({\bf
    r}_{l+1} - {\bf r}_l) P^\alpha_{Str}({\bf r}_{l} - {\bf r}_{l-1})
  \cos(\pi^\alpha_l).  \label{HQ}
\end{equation}
\vskip -0.8cm

\vskip -0.3cm
The above string model is invariant under rotation of the string in space. As 
will be discussed below, we find that for physical choices of the parameters 
the invariance under symmetry operations of the lattice is broken. The string 
acquires a sense of {\em direction} in space. This occurs even when the string 
is critical (delocalized in space). The string's trajectories, on average, 
are such that they move forward in one direction while the string might 
delocalize in the other direction. 

The relation of the string problem to surface models is established by using 
the Suzuki-Trotter mapping which maps a 2D quantum problem to a 
2+1 D classical problem. A classical model of two coupled RSOS 
(restricted solid-on-solid) surfaces results. These are classical models 
for surface roughening\cite{dennijsrev} where overhangs are not allowed. 
For the quantum string case, the two RSOS surfaces describe the motion 
of the string in the $x$ and $y$ spatial directions. 

Skipping detailed calculation, the partition function of the quantum string 
can be mapped to the following classical problem\cite{eskes1},
\begin{eqnarray}
  && {\cal Z} = \lim_{n \rightarrow \infty} Tr e^{ {\cal H}_{eff} }
  \nonumber \\ &&{\cal H}_{eff} = \sum_{l,k} \left[ { {\cal K} \over
    n} \delta( |x_{l+1,k}-x_{l,k}| - 1 ) \delta( |y_{l+1,k}-y_{l,k}| -
  1 ) \right.  \nonumber \\ && + \sum\limits_{i,j = 1}^{2} {{\cal
      L}_{ij} \over n} \delta( | x_{l+1,k} - x_{l-1,k} | - i) \delta(
  | y_{l+1,k} - y_{l-1,k} | - j ) \nonumber \\ && + \ln ( {{\cal T} 
  \over n} ) \left[ \delta ( | x_{l,k+1} - x_{l,k} | -1 ) + \delta 
   ( | y_{l,k+1} - y_{l,k} | -1 ) \right].
\label{Heff}
\end{eqnarray}
where $k$ is the trotter index and with the constraint $| x_{l,k+1} - 
x_{l,k} |\leq 1$ and $| y_{l,k+1} - y_{l,k} | \leq 1$. The above 
classical model can be viewed as two coupled RSOS surfaces, $x_{l,k}$ and 
$y_{l,k}$.  The $x$ coordinate of particle $l$ at the trotter height $k$ is 
the height at position $(l,k)$ in the first surface and similarly
the $y$ coordinates define a second RSOS surface,
coupled strongly to the first by the above classical interactions.


Let us first discuss the numerical results. It is clear that the 
directedness property is a {\em global} quantity. For a string living in 
2D lattice with open boundary conditions, directedness means that if it 
start at, say, the left boundary it has to end at the right boundary and 
will never end at the top or the bottom boundaries of the lattice. Although 
in the above model one can introduce a local order parameter to measure 
the directedness of a string, a more general quantitative measure for this 
global property can be constructed. This measure is not easily evaluated 
analytically but it can easily be calculated numerically; most importantly 
it illustrates clearly and effectively the directedness phenomenon. Every 
string configuration $s$ defines a curve in the 2D space [$x(t),y(t)$], where 
$t$ could for instance be the discrete label of the successive particles 
along the string. When this curve can be parametrized by a {\em single-valued} 
function $x(y)$ or $y(x)$, we call the string configuration directed. The 
quantum string vacuum is a linear superposition of many string configurations. 
When all configurations in the vacuum correspond to single valued functions
$x(y)$ or $y(x)$, the string vacuum is directed. At zero temperature, 
the ground state wave function of the string is 
$ | \Psi_0 \rangle = \sum_{\{ x_l, y_l \} } \alpha_0 ( \{ x_l, y_l \})
  | \{ x_l, y_l \} \rangle~, $
where every state in string configuration space ($| \{ x_l, y_l \}
\rangle$) corresponds to a trajectory [$x(t),y(t)$]. Consider first
the case of a continuous string. For every configuration, the total
string arclength is given by
\begin{equation} \label{totl}
  L(\{x_l,y_l\})_{tot}= \int ds = \int \sqrt{dx^2+dy^2}~.
\end{equation} 

\vskip -0.7cm
\begin{figure}
\hspace{12ex}\epsfig{figure=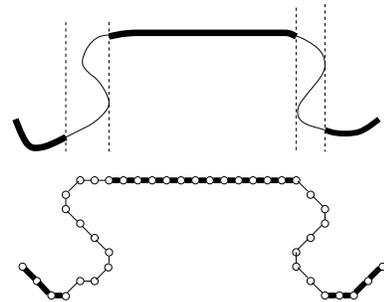,width=5cm}   
\caption{ {\footnotesize Measuring the directedness in the continuum 
case (a) and on the lattice (b). The heavy lines correspond to the directed 
part.}}
\label{directeddef}
\end{figure}

\vskip -0.3cm
Consider now an indicator function $g_y(x)$ which equals 1 when
the string is single-valued when projected onto the $x$-axis, and zero
otherwise, and analogously a function $g_x(y)$ for the $y$-axis, (see 
Fig. \ref{directeddef}). 
The total directed lengths in the $x$ and $y$ directions are
defined as
\begin{eqnarray} \nonumber
  L(\{ x_l, y_l \})_{dir, x} & = & \int dx ~ g_y(x) \sqrt{1 + \left(
    {{dy}\over{dx}}\right)^{2} }~,\\ L(\{ x_l, y_l \})_{dir, y} & = &
  \int dy ~ g_x(y) \sqrt{1 + \left( {{dx}\over{dy}}\right)^{2}
    }~.\label{totdx}
\end{eqnarray}
The measure of directedness is then defined as
the larger of $N_{dir}^{x} (0)$ and $N_{dir}^y (0)$, where
\begin{equation} \label{dirden}
  N_{dir}^{\eta} ( 0 ) = \sum_{ \{ x_l, y_l \} } | \alpha_0 ( \{ x_l,
  y_l \} )|^2 { {L(\{ x_l, y_l \})_{dir, \eta}} \over { L({x_l,
        y_l})_{tot} } }~,
\end{equation}
and $\eta=x,y$. On the lattice, one measures the directedness in analogy 
with the above definition, except that we just count the number of 
directed bonds, irrespective of whether they are oriented diagonally or 
horizontally. The finite temperature measure of the directedness density 
is simply given by thermally averaging the above definition.
\begin{equation} \label{dirdenT}
  N^{\eta}_{dir} ( T ) = \sum_{n} e^{ -\beta (E_n - E_0) }
  N_{dir}^{\eta} ( n )~,
\end{equation}  
\vskip -0.3cm
\noindent where $N_{dir}^{\eta} ( n )$ is the directedness density of an 
excited string with energy $E_n$. 

To study the directedness property we performed exact diagonalization and 
quantum Monte-Carlo studies. Although the quantum Monte-Carlo study is the 
more extensive we start by discussing the exact diagonalization results, as 
it will give a clear indication for the symmetry breaking directly at zero 
temperature. Here we consider an $N \times N$ lattice. We think of a string 
living in such a finite lattice as part of an infinite one and therefore the 
ends of the string should live on the boundaries of the cluster. To fix the 
length of the string inside the cluster, we take as a criterion that the 
energy per particle be minimum. We plot the energy per particle versus the 
number of particles in the string. The minimum defines the optimal length of 
a string in the cluster. Upon setting the parameter ${\cal L}_{11}$ to zero 
and investigating different points in the parameter space, we found that the 
optimal length one should consider is the linear dimension of the lattice. 
Therefore in an $N \times N$ lattice we will consider a string of length $N$. 
Such a string 
can be directed along the $x$ (horizontal) or $y$ (vertical) direction. If the 
directedness assumption is fulfilled, the Hilbert space will effectively split 
into two subspaces: strings directed along the $x$ direction and those along 
the $y$. If nondirected strings are present there should be a non-zero 
tunnelling probability between the two sectors. By measuring the probability 
to tunnel from the $x$- to the $y$- sectors as a function of the linear 
dimension of the system, it should be possible to see the tendency towards 
spontaneous directedness symmetry breaking in the thermodynamic limit. Table  
1 gives this tunnelling probability for different 
points in the parameter space. For all cases we set ${\cal L}_{11} = 0$. 
The choice of these points was motivated by the directed string 
problem\cite{eskes1}. The data are shown for lattices up to $7\times 7$. For a 
$9\times 9$ lattice the tunnelling probability turns out to be less than the 
accuracy of our numerical technique.

\vskip 0.1cm
\noindent (a)
\begin{tabular}{|l|c|}
\hline 
Lat. & Prob. \\ 
\hline  
$3\times 3$ & $\sim 10^{-1}$ \\ \hline
$5\times 5$ & $\sim 10^{-5}$\\ \hline
$7\times 7$ &$\sim  10^{-11}$\\ \hline
\end{tabular}
(b)
\begin{tabular}{|l|c|}
\hline 
Lat. & Prob. \\ 
\hline  
$3\times 3$ & $\sim  10^{-1}$ \\ \hline
$5\times 5$ & $\sim 10^{-3}$\\ \hline
$7\times 7$ &$\sim  10^{-9}$\\ \hline
\end{tabular}
\noindent (c)
\begin{tabular}{|l|c|}
\hline Lat. & Prob. \\ \hline 
$3\times 3$ & $\sim 10^{-2}$ \\ \hline
$5\times 5$ & $\sim 10^{-5}$\\ \hline
$7\times 7$ &$\sim  10^{-10}$\\ \hline
\end{tabular} 

\vskip 0.1cm
(d)
\begin{tabular}{|l|c|}
\hline 
Lat. & Prob. \\ 
\hline 
$3\times 3$ & $\sim  10^{-2}$ \\ \hline
$5\times 5$ & $\sim 10^{-4}$\\ \hline
$7\times 7$ &$\sim  10^{-6}$\\ \hline
\end{tabular} 
\hskip 0.3cm
\hspace{10ex} (e)
\begin{tabular}{|l|c|}
\hline Lat. & Prob. \\ \hline  
$3\times 3$ & $\sim  10^{-1}$ \\ \hline
$5\times 5$ & $\sim 10^{-4}$\\ \hline
$7\times 7$ &$\sim  10^{-7}$\\ \hline
\end{tabular}

\noindent Table 1:{\footnotesize Tunnelling probability at different points 
in the parameter space (${\cal K}, {\cal L}_{12}, {\cal L}_{22}$) (a) (0,0,0), 
(b) (0, -0.25, -1.0), (c) (6.0, -3.0, -2.0), (d) (7.0, -4.0, -6.0), (e) 
(3.0, -3.25, -3.0). }

\vskip 0.1cm
These results clearly indicate that in the thermodynamic limit there is no 
tunnelling between the two sectors and the string should be directed either 
along the $x$ or the $y$ direction.

\begin{figure}
\hspace{8ex}\epsfig{figure=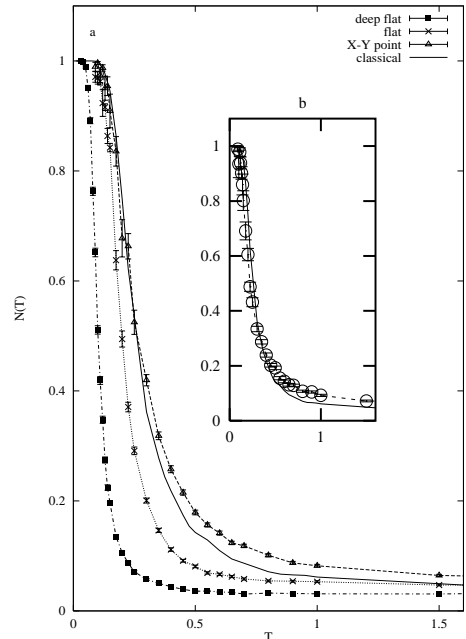,width=6cm}   
\caption{ {\footnotesize Monte-Carlo results for the directedness measure 
$N_{dir} ( T )$ at 4 points in parameter space. (see the text). }}
\label{fig_dir_ord}
\end{figure}

We then used quantum Monte-Carlo to calculate the directedness density as 
a function of temperature, $N_{dir}(T)$ (Eq.\ \ref{dirdenT}).
Our results are displayed in Fig.\ \ref{fig_dir_ord}. Four points in 
parameter space were considered. These points are representative for phases 
with a varying strength of the quantum fluctuations and serve to substantiate 
our conclusion. In (a), the dashed line is the result when all classical 
energies are zero, i.e. for optimally quantum string. The dashed-dotted and 
dotted lines correspond to all potential energy parameters set to zero except 
that ${\cal K} = 4.0$ and $1.8$, respectively, corresponding to a string 
localized in the (1,0) or (0,1) directions. Decreasing the K parameter causes 
stronger local fluctuations. The full line is the result for a 
classical string (${\cal T}=0$) where only flat segments and ${\pi/2}$ corners 
are allowed (no diagonal segments). The same classical result is shown again 
in Fig.\ \ref{fig_dir_ord}(b) together with the result at the point [${\cal K} 
= 0.5$, ${\cal L}_{21} = -0.25$, ${\cal L}_{22} = -1.0$, ${\cal L}_{11} = 0$] 
corresponding to a free (critical) string. 

\vskip -0.4cm

\vskip -0.3cm
The classical string (${\cal T}=0$) would be flat at zero temperature, 
directed along (say) a (1,0)
direction. A local `corner' (Fig.\ \ref{fig_energies}{\em a}) would be an 
excitation with energy ${\cal L}_{11}$ (alternatively, one could consider 
two kinks). Clearly, a single corner suffices to destroy the directedness 
of the classical ground state. At any finite temperature, the probability 
of the occurrence of at least one corner is finite: $P = N \exp ( - \beta 
{\cal L}_{11})$. Hence, directedness order cannot exist at any non-zero 
temperature, for the same reason that any long range order is destroyed at 
any finite temperature in one dimension. In the simulations the string is of 
finite length, and the 
infinite temperature limit of $N_{dir} ( T )$ is therefore
not zero but rather a 
small but nonzero value\cite{selfavoid} ($\sim 0.03$ for a domain wall 
of length 50).  $N_{dir} ( T )$ is already close to this value for all 
temperatures of order ${\cal L}_{11}$ and larger. For an infinitely long 
domain wall $N_ {dir}( T )$ drops very fast to zero with increasing 
temperature. For low $T$ where $T \ll {\cal L}_{11}$, $N_{dir} ( T )$ grows 
rapidly to 1. Again, because the string is of finite 
length, it becomes directed already at a finite temperature: For all 
temperatures such that $L\exp(-{\beta}{\cal L}_{11}) < 1$ the string 
configurations in our simulations are typically completely directed. An 
infinitely long classical string becomes directed only at $T = 0$, of 
course, since at any nonzero temperature always some corners will occur 
in a sufficiently long string. 

The results for the quantum string look always similar to the classical one. 
For temperatures higher than the kinetic scale, $T \gg {\cal T}$, all curves 
approach each other and the classical limit is reached. At low $T$, 
$T \ll {\cal T}$, $N_{dir} ( T )$ grows rapidly to 1. As in the 
classical case, it reaches this value at a finite temperature for a 
finite length string. This is even valid for the pure quantum string, 
where all classical energies are zero (dashed line in Fig.\ 
\ref{fig_dir_ord}(a)). Again this can be understood in terms of an effective 
corner or bend energy ${\bar {\cal L}}$ that is produced by the quantum 
fluctuations. In analogy to the classical case, the probability for the 
occurrence of a bend is proportional to $\sim \exp(- \beta {\bar {\cal L}})$.  
At zero temperature no bend is present and the string becomes directed.
 A finite length string 
effectively becomes directed already at a temperature such that 
$L \exp ( - \beta{\bar {\cal L}}) < 1$. At intermediate temperatures, 
where the temperature is of the order of the kinetic term, the situation is 
less clear. Especially in this region, all the various energies may play a
role, and the interplay of these on the directedness is rather complicated.
Nevertheless, as is clear from the data of Fig.\ \ref{fig_dir_ord}(a),
this region connects the high and low temperature limits smoothly.
Finally, by comparing the results for the three quantum strings in this 
figure it is also clear that when the string is more quantum mechanical 
$ N_{dir} ( T )$ is higher.

\begin{figure}
\hspace{8ex}\epsfig{figure=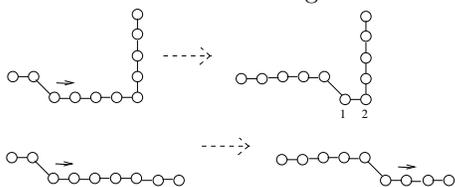,width=6cm}   
\caption{{\footnotesize Bends blocks the propagation of links along the 
string. Note holes 1 and 2 cannot move.}}
\label{fig_bends}
\end{figure}

The spontaneous directedness of the quantum string for ${\cal L}_{11} 
\ge 0$ can be understood by the following argument.  The $\pi/2$ bends 
in strings block the propagation of links along the chain.  Close to 
the bend itself the
particles in the chain cannot move as freely as in the rest of the
chain. This effect is shown in Fig.\ \ref{fig_bends}.

\vskip -0.4cm
 In space-time
the $\pi/2$ bend is like a straight rod in time.  Therefore the
presence of such kinks increase the kinetic energy.  For the argument
it makes little difference whether the bend consists of a single
$\pi/2$ corner or two $\pi/4$ corners.  This confirms that {\it it is
  the kinetic energy which keeps the strings oriented along one
  particular direction}.  In terms of a directedness order parameter
this result implies that such a quantity is always finite, except when
${\cal L}_{11} \ll 0$ or when the hopping term vanishes (it is easy to
see that in the classical case, ${\cal T}=0$, in many regions of
parameter space the problem becomes that of a self-avoiding walk on a
lattice in the limit $T\rightarrow 0$). For the two equivalent RSOS
surfaces this means that one of the two surfaces spontaneously orders 
while the other RSOS sheet can be either ordered or disordered.

Our general conclusion, based also on Monte-Carlo studies of the 
behavior in many other points in the parameter space, is that {\it 
apart from some extreme classical limits, 
the general lattice string model at zero temperatures is a directed string}. 
The qualitative picture of ${\pi /2}$ corners blocking the propagation of 
kinks appears to be a natural explanation for these numerical findings.

{\em Acknowledgements.} 
 We are grateful ot Henk Eskes for a collaboration of which this work 
is an outgrowth.

\end{multicols}

\end{document}